\documentclass[final]{svjour3}
\usepackage{graphicx}
\usepackage{float}
\usepackage{rotating}
\usepackage{amssymb}
\usepackage{siunitx}
\usepackage{array}
\usepackage{cite}
\usepackage{upgreek}
\newcolumntype{M}[1]{>{\centering\arraybackslash}m{#1}}
\DeclareSIUnit\sq{\ensuremath{\Box}}
\usepackage{mathptmx}
\usepackage[numbers,sort]{natbib}
\usepackage[colorlinks]{hyperref}

\makeatletter

\journalname{Journal of Low Temperature Physics}

\bibpunct{[}{]}{,}{n}{}{,}

\begin{document}

\newcommand{\hdblarrow}{H\makebox[0.9ex][l]{$\downdownarrows$}-}
\title{A Thermal Kinetic Inductance Detector Pixel Design for CMB Polarization Observations at 90/150~GHz bands}

\author{Ye Chai$^{1,2}$ \and Shibo Shu$^{2\dagger}$ \and Yongping Li$^{2}$ \and Jiamin Sun$^{3}$ \and Zhouhui Liu$^{2}$ \and Yu Xu$^{2}$ \and Daikang Yan$^{2}$ \and Zhengwei Li$^{2}$ \and Yang Liu$^{2}$ \and Yiwen Wang$^{4}$ \and Weijie Guo$^{5}$\and Juexian Cao$^{1}$ \and Congzhan Liu$^{2}$}

\institute{$^1$ Xiangtan University, China\\
$^2$ Institute of High Energy Physics, Chinese Academy of Sciences, China\\
$^3$ Shandong Institute of Advanced Technology, China\\
$^4$ Southwest Jiaotong University, China\\
$^5$ International Quantum Academy, Shenzhen 518048, China\\
$^\dagger$\email{shusb@ihep.ac.cn}}

\maketitle

\begin{abstract}
The highly sensitive millimeter-wave telescope is an important tool for accurate measurement of Cosmic Microwave Background (CMB) radiation, and its core component is a detector array located in a cryogenic focal plane. The feasibility of utilizing thermal kinetic inductance detectors (TKIDs) for CMB observations has been demonstrated. We propose a pixel design of TKIDs for observing CMB through atmospheric windows for observations in the 90/150~GHz bands. The TKIDs are designed to achieve photon noise limited sensitivity with overall noise equivalent power can be less than 20~aW$/\sqrt\mathrm{Hz}$. Silicon-rich silicon nitride is used as the dielectric of TKIDs, and Al is used as the inductance material. Two pairs of probes are designed on the pixel to divide the signal into two polarization directions. Orthogonal transducer and diplexer are used for signal conversion and frequency division. Assuming lossless dielectric, the coupling efficiency of a single pixel is around 90\%. This pixel design will be utilized for future large-scale TKIDs array designs for CMB observations.


\end{abstract}

\section{Introduction} 

Cosmic microwave background (CMB) radiation is an important target for studying the early universe. The tensor-to-scalar ratio ($r$) of primordial modes during inflationary epoch can be constrained by the B-mode polarization of CMB.\cite{ade2014planck, penzias1979measurement, abazajian2016cmb, collaboration2014experiment, hui2018bicep}. In order to constrain $r$, it is necessary to deploy a large number of highly sensitive detectors to observe CMB at different frequency bands. Currently, transition edge sensors (TESs) have been widely utilized in research experiments investigating the polarization of the CMB\cite{irwin99transition, zhang2020characterizing}. However, due to the requirement of using superconducting quantum interference devices (SQUIDs) amplifier for TESs readout, a large number of TESs array need to be combined with the corresponding SQUIDs amplifier readout array, which increases the complexity of the detector system\cite{clarke2004squid}. As an alternative detector technology, kinetic inductance detectors (KIDs) simplify the integration of the system by using microwave frequency division multiplexing technology, which does not rely on additional readout chips\cite{Day:2003a}. KIDs use superconductor inductors as photon absorbers. Therefore, the detection frequency range is limited by the transition temperature ($T_c$) of the superconducting material. As a commonly used superconducting material, aluminum has a $T_c$ of 1.2~K, which means it is unable to absorb photons below $73~\mathrm{GHz/K}\times T_c\approx 90~\mathrm{GHz}$. Although Al/Ti bilayer\cite{Catalano2015Bilayer} and AlMn\cite{Jones2017AlMnKID} have lower $T_c$ than Al, their low $T_c$ also requires a lower operating temperature. This usually results in a high two-level system (TLS) noise and makes it difficult to work at $>200$~mK in a He sorption fridge, like BICEP Array\cite{ade2014detection} and AliCPT1\cite{salatino2020design}.

Thermal kinetic inductance detector (TKID)\cite{steinbach2018thermal, wandui2020thermal} is a variation of KID, and can also use the readout technology of microwave frequency division multiplexing. In TKIDs, the optical power is thermalized on the thermal island, causing the island temperature to rise. This process changes the quasiparticle density in the inductor, and then changes the inductance, and finally causes the resonant frequency to change. The power of photons can be detected by tracking the change of island temperature by monitoring the change of resonant frequency. Through this process, the optical absorption is no longer limited by inductor $T_c$, allowing us to detect 90~GHz or even 40~GHz band with Al. The high $T_c$ of Al allows us to detect at a high operating temperature $> 200$~mK, which can decrease TLS noise. Based on this, we present a dual-polarization TKID pixel design for CMB B-mode observations at 90~GHz and 150~GHz bands in this paper. 

\section{Single pixel design}

\begin{figure}
    \centering
    \includegraphics[width=0.95\textwidth]{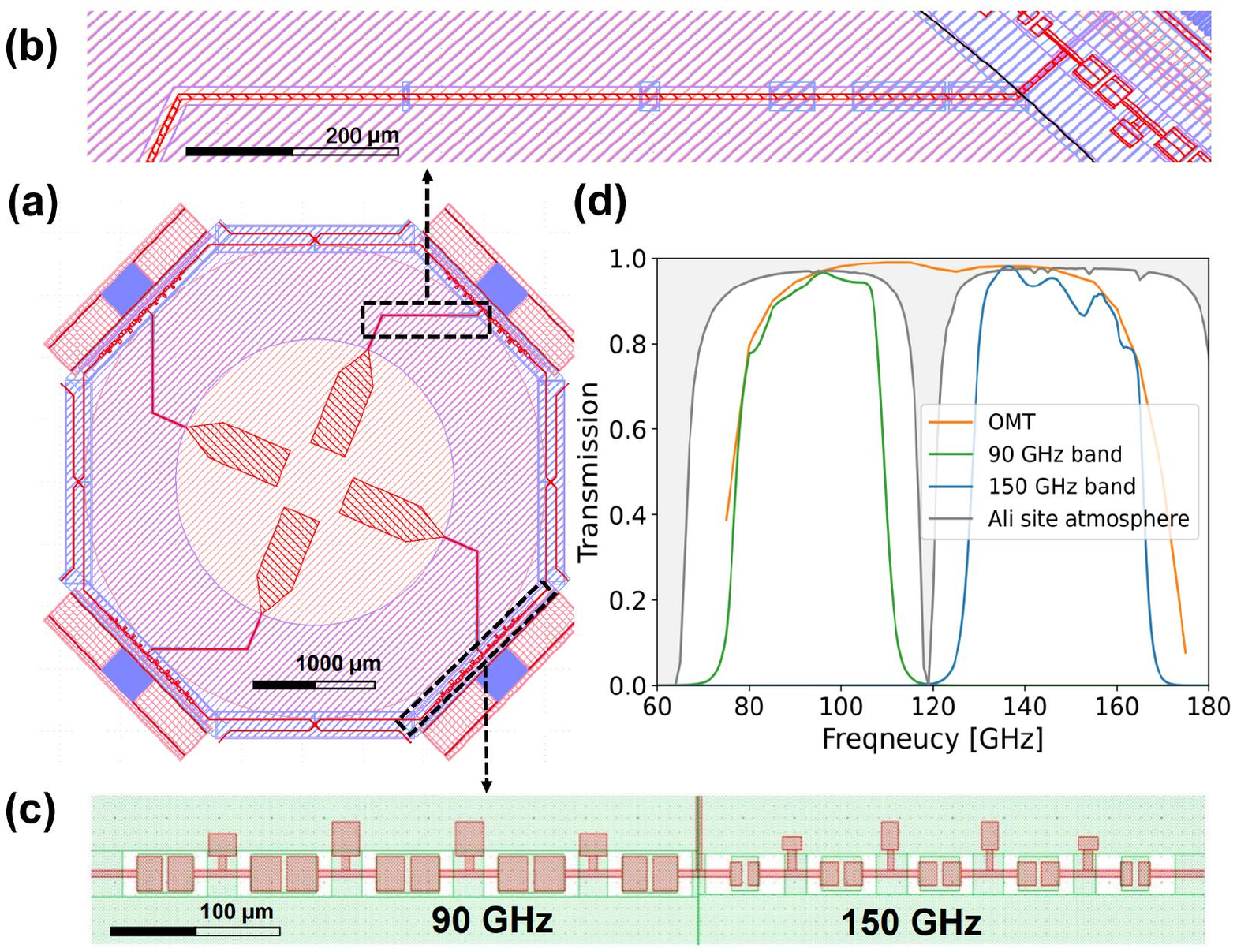}
    \caption{(a)Mask design of a single pixel. (b)The simulation model constructed within Ansys HFSS. (c)Design of the diplexers. (d)Assuming the transmission efficiency of consider OMT+CPW-to-MS+DPX+crossunder under lossless.}
    \label{fig:1}
\end{figure}

CMB signal peaks at millimeter-wave. Ground-based observations are usually performed at 90~GHz and 150~GHz atmospheric windows around the CMB peak. 
We designed a single pixel structure for CMB polarization signals in these two bands. The single pixel contains four TKIDs, two for collecting signals from 90~GHz bands and two for collecting signals from 150~GHz bands. In the readout bandwidth of 4-8~GHz, the frequency interval of the adjacent detectors is designed to be 5 times their bandwidth to minimize cross talk. We also designed an orthomode transducer (OMT), lumped-element diplexer, cross-under structure, and lumped resistor to realize the signal conversion and transmission in the single pixel structure. We used Sonnet EM to simulate filters and TKID,and Ansys HFSS module to simulate OMT, diplexer, cross-under, and lumped resistor.
In our design, horn antennas are used for a broadband detection. These antennas can be either machined from metal or stacked using silicon wafers~\cite{Simon:2018feedhorn}. 
To convert waveguide signal onto planar transmission lines, a planar 4-probe OMT are inserted into the circular waveguide with a backshort located at a quarter wavelength behind the probes~\cite{Grimes:2007a, McMahon:2012a, Shu:2016a}, as shown in Fig.~\ref{fig:1}a. We choose spline-profiled horns using stacked lithography-patterned Si wafers to maximize the number of pixels. The diameter of the circular waveguide is 2.3~mm in design with a cut-off frequency of 76.4~GHz. 

A silicon wafer with 200~nm-thick silicon oxide and 1000~nm silicon nitride is used. The geometry of the probes together with the 136.5~$\mathrm{\Omega}$ coplanar waveguide are optimized to match the high-impedance waveguide mode. An impedance transition is designed to match the 7.5~$\mathrm{\Omega}$ microstrip line (see Fig.~\ref{fig:1}b). A lumped-element diplexer is designed with a $<1\%$ frequency overlap due to a fast roll-off~\cite{Shu:2022multi}, shown in Fig.~\ref{fig:1}c. This design also has no higher-order passband, and the size is decreased by a factor of 2 compared with stub filters, which will decrease the attenuation in transmission line. The OMT probes divide the signal in the same polarization into two parts along the two probes. An intersection structure is needed to add these two parts up, as each polarization signal is separated into two bands. The microstrip lines input the same polarization signal into the same TKID, so the microstrip lines will cross. In order to prevent short circuit and reduce crosstalk, we designed a cross-under structure. The cross-under structure is designed with power leakage smaller than -27~dB. Finally, the two signal from the same polarization but with $180^{\circ}$ phase difference is converted to heat on a lumped resistor~\cite{Westbrook:2016pb}. This thermal interface makes the design compatible for both TKIDs and transition edge sensors. The final transmission band assuming lossless dielectric is shown in Fig.~\ref{fig:1}d. The simulated bands are 77-109~GHz and 128-166~GHz.

\section{Thermal kinetic inductance detectors design}

\begin{figure}
    \centering
    \includegraphics[width=0.95\textwidth]{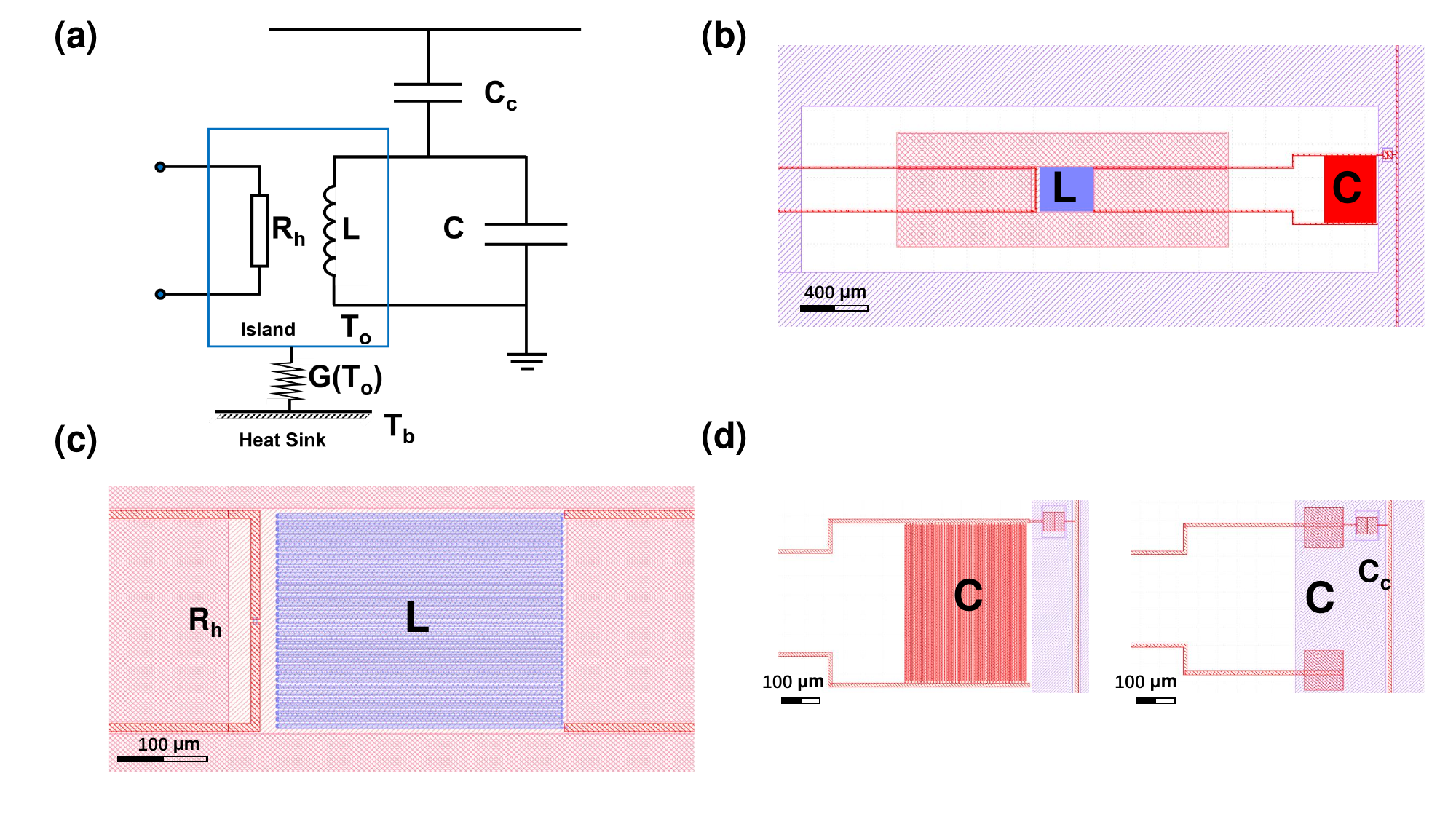}
    \caption{(a)The circuit model of a TKID, thermal island circuit is shown in blue box. (b)Mask design of a TKID. (c)The inductor $L$ and the gold resistor $R_h$ in a TKID. (d)The capacitor and the coupling capacitor in a TKID.}
    \label{fig:2}
\end{figure}

Fig.~\ref{fig:2}a shows the circuit model of our lumped-element TKID design. The meandered inductor is made from 200~nm thick Al, which has a $T_c$ of 1.2~K, a sheet resistance of 0.0208~Ohm$/\ensuremath{\Box}$, and a kinetic inductance of 0.075~pH$/\ensuremath{\Box}$. Both the inductor width and spacing are 2~$\mathrm{\mu m}$. The optimized inductor length and volume are 21.19~mm and 8.476$\times10^3$~$\mathrm{\mu m}^3$, respectively. The resonant frequency $f_r$ is tuned by the capacitor size (see Fig.~\ref{fig:2}b). The wafer has a 1~$\mathrm{\mu m}$ thick silicon nitride and a 200~nm thick silicon oxide on top of silicon substrate, so both interdigital capacitor and parallel plate capacitor are used in the test device for measuring the TLS noise from both the wafer and the 300~nm-thick silicon-rich silicon nitride ($\mathrm{SiN_x}$) used for the capacitors (see Fig.~\ref{fig:2}d).

The resonator is coupled to a 50~$\mathrm{\Omega}$ coplanar waveguide feedline through a coupling capacitor $C_c$. The trace of the feedline is on top of the $\mathrm{SiN_x}$ layer and the ground plane is at the bottom. This design provides a direct connection from the feedline to the TKID without using air-bridges. All feedlines and capacitor structures are designed with 400~nm thick niobium ($T_c\sim \mathrm{9.2~K}$), ensuring that the thermal response is all from the aluminum inductor. In our test device, a gold resistor ($R_h$) is designed to generate thermal loading on the thermal island for dark testing (see Fig.~\ref{fig:2}a). The $R_h$ is designed to be 8~$\mathrm{\mu m}$ long with 4~$\mathrm{\mu m}$ wide (see Fig.~\ref{fig:2}c), and 50~nm thick, which has a resistance of 1.45~Ohm according to our measurements. $R_h$ and inductor together form the thermal island, which is released using an etching process. The thermal island is anchored to the wafer by four legs each 10~$\mathrm{\mu m}$ wide. These legs act as weak thermal link from the thermal island to the wafer. The legs will also be fabricated with Ni of the same thickness as the feedlines.

The differential equation:
\begin{equation}
    C\frac{dT}{dt} = -P_{leg} + P_{read} + P_{opt}
\end{equation}
represents the thermal response of TKIDs, where $C$ is the heat capacity of the thermal island. $P_{read}$ is the readout power dissipated by the inductor. $P_{leg}$ is the net heat flow through the thermal link and its model is $P_{leg}=K(T^{n}-T_{bath}^{n})$, where $K$ is a coefficient and $n$ is the power law index. $P_{opt}$ represents the optical loading at detector. We use the atmospheric transmission data collected at 45-degree angle to the horizon for the optical load calculation. The median temperature at 90~GHz bands and 150~GHz bands are 10.6~K and 12.6~K, respectively. The result is 7~pW at 90~GHz bands and 10~pW at 150~GHz bands. Assuming 50\% transmittance in both bands, the final power is set to 3.5~pW at 90~GHz bands and 5~pW at 150~GHz bands, respectively. The corresponding thermal island temperatures under these two loadings are 378~mK and 400~mK, respectively. 378~mK and 400~mK as operating temperature ($T_o$), we can assume $P_{read} \ll P_{opt}$, and $P_{opt} = P_{leg}$ in equilibrium.

\begin{figure}
    \centering
    \includegraphics[width=0.95\textwidth]{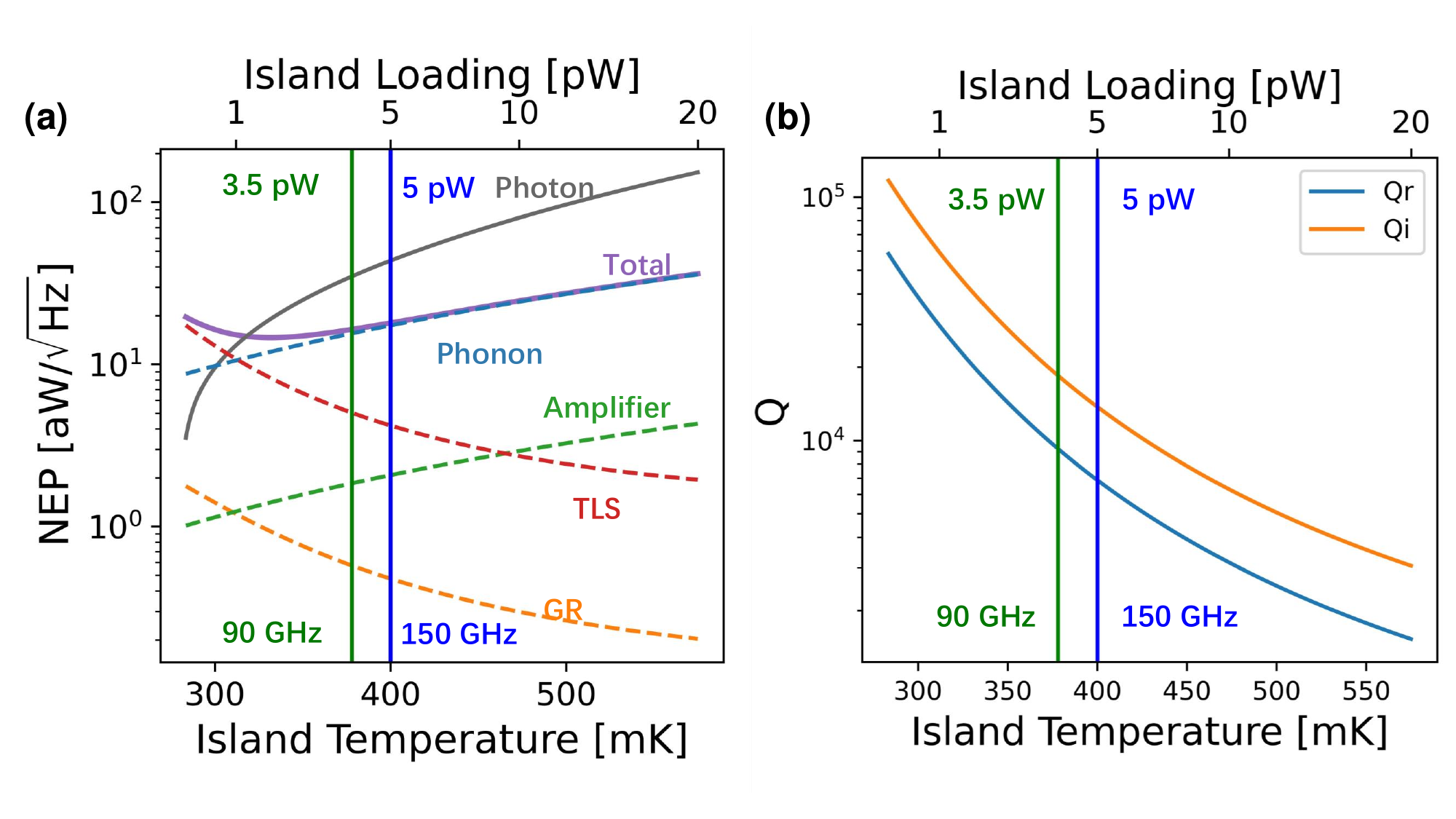}
    \caption{(a)Detector noise model of our TKID as a function of the island temperature showing each noise term, top axis and bottom axis correspond to the relationship between the received photon energy and the thermal island temperature. (b)The relationship between quality factor and island temperature variations.}
    \label{fig:3}
\end{figure}

Noise performance is an important index of detector performance. We perform theoretical calculations of the expected noise equivalent power ($\mathrm{NEP}$) by optimizing the design parameters, as shown in Fig.~\ref{fig:3}a \cite{wandui2020thermal, mccarrick2018design, zmuidzinas2012superconducting}. When calculating the overall NEP ($\mathrm{{NEP}_{total}}$), five noise terms are considered: photon, phonon, generated-recombination, two-level system, and amplifier noise\cite{mather1982bolometer, pirro2017advances, lindeman2014resonator}, among which the relationship is given by:
\begin{equation}
    \mathrm{NEP_{total}^2} = \mathrm{NEP_{phonon}^2} +  \mathrm{NEP_{gr}^2} + \mathrm{NEP_{TLS}^2} + \mathrm{NEP_{amp}^2}
\end{equation}

When the phonon noise in the thermal equilibrium state of the system is not considered, the phonon noise generated by the thermal response is due to the randomness in the energy transfer along the thermal link. The $\mathrm{NEP_{phonon}}$ is determined by the thermal conductance $G(T_o)$ as follows: $\mathrm{NEP_{phonon}^2}$ = $4F(T_o, T_{bath}){k_B}{T_o^2}{G(T_o)}$, where $F(T_o, T_{bath})$ represents the temperature gradient along the thermal link, and $k_B$ is the Boltzmann constant.The $\mathrm{NEP_{phonon}}$ can be simplified to $\mathrm{NEP_{phonon}^2} = 4\tilde{F}(T_o, T_{bath}){k_B}{T_o}{P_{opt}}$, where $\tilde{F}(T_o, T_{bath})$ = $F(T_o, T_{bath})\cdot{n}/(1 - ({T_{bath}/T_o})^n)$\cite{wandui2020thermal}. In the designed device, we use $G(T_o) = 28.54$~pW/K, $F(T_o, T_{bath}) = 0.6$, $\tilde{F}(T_o, T_{bath}) = 2.73$. At the thermal island loading of  3.5~pW, $\mathrm{NEP_{phonon}}$ = 15.57~aW$/\sqrt\mathrm{Hz}$ and $\mathrm{NEP_{photon}}$ = 35.69~aW$/\sqrt\mathrm{Hz}$, respectively. When the thermal island loading of  5~pW, $\mathrm{NEP_{phonon}}$ = 17.38~aW$/\sqrt\mathrm{Hz}$ and $\mathrm{NEP_{photon}}$ = 43.82~aW$/\sqrt\mathrm{Hz}$, respectively. Under both loadings, phonon noise is much smaller than photon noise. 

The generation-recombination noise, arising from the random processes of quasiparticle generation and recombination in TKIDs, can be simulated using non-equilibrium statistical mechanics\cite{zmuidzinas2012superconducting, wilson2004quasiparticle}. At 90~GHz bands and 150~GHz bands, the $\mathrm{NEP_{gr}}$ for both is less than 1~aW$/\sqrt\mathrm{Hz}$, which is negligible.  TLS noise is a kind of noise term brought by dielectric. There is no corresponding microscopic at present. In the verification experiment, we measured the TLS noise spectrum of 300~nm thick $\mathrm{SiN_x}$ at different temperatures, which provided data for us to calculate $\mathrm{{NEP}_{TLS}}$. When considering the amplifier noise, the noise temperature of the amplifier is set to 5~K. It is assumed that the coupling quality factor is equal to the internal quality factor ($Q_i$) in the case of optimal coupling\cite{zmuidzinas2012superconducting, khalil2012analysis}. 
\begin{center}
 \begin{table}[H]
 \centering
 \caption{NEP values of each noise term in TKIDs at 90~GHz bands and 150~GHz bands}
 \renewcommand{\arraystretch}{1.7}
 \begin{tabular}{ p{1.5cm}<{\centering} p{0.7cm}<{\centering} p{0.9cm}<{\centering} p{0.9cm}<{\centering} p{0.9cm}<{\centering} p{0.9cm}<{\centering} p{0.9cm}<{\centering} p{0.9cm}<{\centering} p{0.9cm}<{\centering}}
 \hline
 Frequency band & $P_{opt}$ & $\mathrm{NEP_{photon}}$ & $\mathrm{NEP_{phonon}}$ & $\mathrm{NEP_{gr}}$ & $\mathrm{NEP_{TLS}}$ & $\mathrm{NEP_{amp}}$ & $Q_i$ & $Q_r$\\
 \hline
 90~GHz & 3.5~pW & 35.69 & 15.57 & 0.58 & 5.11 & 1.86 & 1.9$\times10^4$ & 0.9$\times10^4$\\
 \hline
 150~GHz & 5~pW & 43.82 & 17.38 & 0.48 & 4.42 & 2.09 & 1.4$\times10^4$ & 0.7$\times10^4$\\
 \hline
 \label{tab:1}
 \end{tabular}
\vspace{-6.5em}
\end{table}
\end{center}
In Fig.~\ref{fig:3}a, each curve represents the variation trend of $\mathrm{NEP}$ of each noise term with the island temperature, with the increase of temperature, $\mathrm{{NEP}_{gr}}$ and $\mathrm{{NEP}_{TLS}}$ are suppressed to a certain extent, and $\mathrm{{NEP}_{amp}}$ will increase. The green and blue lines represent the thermal island temperatures of our devices at 90~GHz bands and 150~GHz bands, respectively. The $\mathrm{NEP}$ of each noise term in TKIDs at 90~GHz bands and 150~GHz bands are listed in Table.~\ref{tab:1}. As can be seen from Fig.~\ref{fig:3}a and Table.~\ref{tab:1}, the detector noise we designed is dominated by photon noise, rather than other noise terms when the loading is above 1~pW. As shown in Fig.~\ref{fig:3}b the resonator quality factor ($Q_r$) and $Q_i$ are limited by island temperature. $Q_r$ and $Q_i$ decrease as the island temperature increases. The $Q_r$ and $Q_i$ for the two island temperatures are also given in Table.~\ref{tab:1}. The $Q_r$ in these two bands are sufficient for our detectors. 

\section{Conclusion}

In this paper, 
we have shown a single pixel design and a thermal kinetic inductance detector design. The optimized pixel module can achieve the coupling efficiency of the whole feedline above 90\% without loss. Phonon noise in TKIDs is low enough to produce background limiting performance with the $\mathrm{NEP_{total}}$ lower than 20~aW$/\sqrt\mathrm{Hz}$, and $\mathrm{NEP_{TLS}}$ = 5.11~aW$/\sqrt\mathrm{Hz}$. This performance is achieved with an operating temperature of 378~mK and an optical power of 3.5~pW. When optical power of 5~pW and operating temperature of 400~mK, $\mathrm{NEP_{TLS}}$ = 4.42~aW$/\sqrt\mathrm{Hz}$. The theoretical calculation results show that the designed single pixel module can be applied to 90~GHz bands and 150~GHz bands for CMB observations. This design will also be used in future large-scale arrays.

\noindent \textbf{Acknowledgements}

This work is supported by National Key Research and Development Program of China (Grant No. 2022YFC2205000).

Data sharing not applicable to this article as no datasets were generated or analysed during the current study.

\bibliographystyle{unsrt}
\bibliography{reference}
\end{document}